\documentclass{epl}
\usepackage{amsmath}

\title{Glassy behavior induced by geometrical frustration in 
a hard-core lattice-gas model} 
\shorttitle{Glassy lattice gas}

\author{
Martin Weigt~\inst{1}\thanks{E-mail:
\email{weigt@theorie.physik.uni-goettingen.de}}
\and 
Alexander K. Hartmann~\inst{1}\thanks{E-mail:
\email{hartmann@theorie.physik.uni-goettingen.de}}
}

\institute{
\inst{1} Institute for Theoretical Physics, University of
G\"ottingen - Bunsenstr. 9, D-37073 G\"ottingen, Germany\\
}


\begin{document}

\maketitle

\begin{abstract}
  We introduce a hard-core lattice-gas model on generalized Bethe
  lattices and investigate analytically and numerically its compaction
  behavior. If compactified slowly, the system undergoes a
  first-order crystallization transition. If compactified much
  faster, the system stays in a meta-stable liquid state and undergoes
  a glass transition under further compaction. We show that this
  behavior is induced by geometrical frustration which appears due to
  the existence of short loops in the generalized Bethe lattices. We
  also compare our results to numerical simulations of a
  three-dimensional analog of the model.
\end{abstract}

The last decades have seen a great interest in understanding the
structural glass transition \cite{Go,BoCu}. The latter is given by a
dramatic dynamical slowing down which prevents the systems from
reaching the equilibrium state at experimental time scales. For
glass-forming liquids, e.g., the glass transition is usually
identified by the temperature where the viscosity exceeds $10^{13}$
poise. A central role in theoretical approaches is played by fully
connected spin-glass models with multi-spin interactions
\cite{KiWo,BoCu}. These models in fact show a purely
dynamical transition towards a glassy state followed by a
thermodynamic transition at lower temperature. The dynamical
transition, even if not strictly present in real materials due to
activated processes, is believed to present some similarities to the
structural glass transition.

The analogies between fully connected spin glasses and structural
glasses are, however, of purely phenomenological nature, missing a
microscopic verification. The main differences for spin-glass models
are, that they have multi-spin interaction, which in addition are 
disordered (quenched disorder) and long-ranged (fully connected
models). So, in particular, these models do not have any crystalline
state. 

Recently, some of these points could be cured by considering models
based on Bethe lattices or Husimi trees, which are the analog of
Bethe lattices for multi-spin interactions: On finite length scales,
these lattices have no inhomogeneities. The corresponding models
undergo a crystallization transition which was found to be of first
order in the most interesting cases. By cooling or compactifying
these models relatively fast, the system stays in the meta-stable liquid
(or paramagnetic) phase. In analogy to their fully-connected
counterparts, they undergo a dynamical glass transition at even lower
temperature, which is followed by a thermodynamic transition
\cite{FrMeRiWeZe,BiMe}. The latter cannot be observed in
simulations (or experiments) since the system does not equilibrate
below the dynamical transition. These models still have, however, the
problem of having multi-spin interactions (which, in some cases, are
introduced by local constraints concerning more than two particles, as
seen in \cite{BiMe}). Similar models were also discussed in the
context of compactified granular matter \cite{LeDe}.

Here we present a lattice-gas model with pure two-particle
interactions. Defined on a usual, i.e., locally tree-like Bethe
lattice, the model does neither show a first-order crystallization nor
a glass transition. If, however, the Bethe lattice is generalized to
include short loops, a dynamical behavior reminiscent of the glass
transition in liquids of hard particles or Lennard Jones liquids
\cite{Ko} is found. This glass transition is induced by geometrical
frustration \cite{frust} which prohibits constructing a globally dense
configuration from purely local rules.

As the original Bethe lattice, also the generalized Bethe lattices
studied here are of mean-field nature. In order to compare our results
to more realistic models, we introduce a three-dimensional lattice gas
showing the same kind of geometrical frustration. We study this model
be means of Monte-Carlo (MC) simulations: The behavior of the model
shares many features with the mean-field case, even if they are in
general much less pronounced.

\begin{figure}[htb]
\twofigures[width=0.4\textwidth]{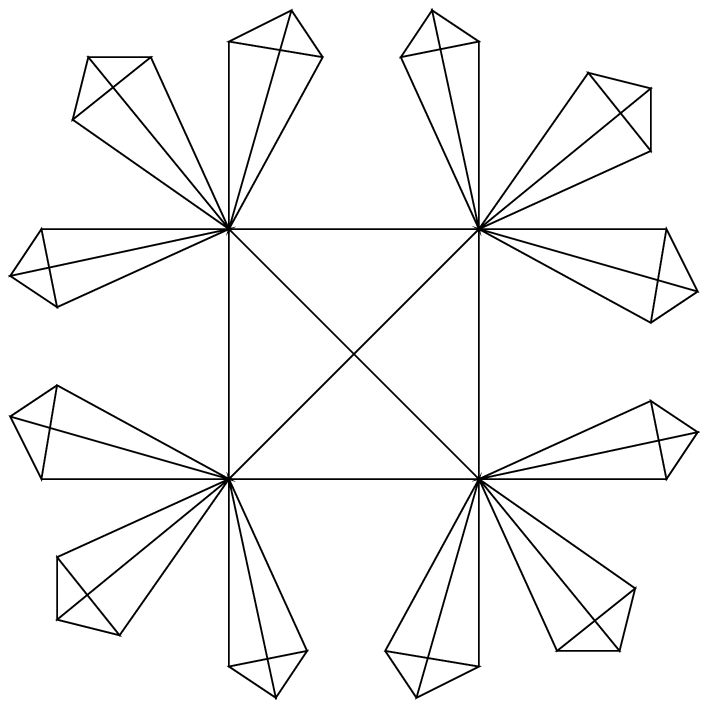}{tetrafix_density.eps}
\caption{ 
  Part of the generalized Bethe lattice. The unit cell is
  equivalent to a clique of $p$ vertices (here $p=4$). Each site is 
  contained in $K+1$ unit cells (here $K=3$).}
\label{fig:bethe}
\caption{ 
  Density $\rho$ as a function of the chemical potential $\mu$, for
  $p=4$, $K=3$. The full lines give analytical results for the liquid
  and the crystalline phases. The spinodal and the crystallization
  points, as well as the dynamic and static glass transition are
  marked by vertical lines (from left to right).  Results are compared
  with numerical compaction curves for random generalized Bethe
  lattices of size $N=999$, averaged over 100 graphs. Here
  $\delta\mu=0.2$ was used.  Inset: Compaction rate dependence for
  high values of $\mu$ for $n_{\rm MC}$ ranging from 5000 (top) to 10
  (bottom).
  }
\label{fig:tetrafixDensity}
\end{figure}

The hard-sphere lattice-gas model can be defined on any graph
$G=(V,E)$. Each site $i\in V$ can be occupied by at most one particle,
$x_i\in\{0,1\}$. Particles are assumed to have a hard core of radius
one, i.e. neighboring sites can never be occupied simultaneously: $x_i
x_j=0$ if $\{i,j\}\in E$. Note that this defines a two-particle
interaction. If the graph is a tree, i.e. if it does not contain any
loops, dense particle packings can be generated from a simple local
rule: Starting with one particle, all neighboring sites have to be
empty due to the hard-core constraint. Particles can again be put to
all second neighbors and so on. Geometrical frustration enters via
loops. If, e.g., some of the second neighbors of a site are directly
connected by an edge, they cannot be occupied simultaneously. The
resulting arbitrarity in selecting occupied second neighbors finally
leads to globally disordered packings of sub-maximal density.

To investigate the influence of geometrical frustration, let us
consider the following generalized Bethe lattices: They are
constructed from basic units consisting of $p$-cliques, i.e.  fully
connected subgraphs of $p$ vertices. From these basic units, a graph
is constructed by joining $K+1$ cliques in every vertex, see
Fig.\ref{fig:bethe}. A usual Bethe lattice is obtained for $p=2$ where
the basic cliques are just simple edges, and the graph is locally
tree-like. For $p>2$, the graph contains triangles, i.e. local loops,
giving rise to geometrical frustration.

The thermodynamic properties of the model are given by the
grand-canonical partition function
\begin{equation}
  \label{eq:xi}
  \Xi = \sum_{\{x_i\}\in \{0,1\}^N} e^{\mu \sum_i x_i} 
  \prod_{\{i,j\}\in E} (1 - x_i x_j) \ .  
\end{equation}
The last product includes the hard-core constraint, it equals one for
allowed packings, and zero if any two neighboring sites are occupied
simultaneously. The chemical potential $\mu$ is coupled to the total
particle number and can be used to tune the particle density. In our
system, we study the compaction of the model which can be achieved by
increasing $\mu$. In particular, densest packings are reached in the
limit $\mu\to\infty$.

The model can be solved most easily using the cavity method
\cite{MePa} which, in the liquid as well as in the crystalline state,
reduces to the Bethe-Peierls iterative method. Note, however, that the
latter is modified with respect to the presence of short loops. On a
coarse-grained level, where every local clique is considered as one
object, the (hyper-)graph becomes locally tree-like and therefore
solvable by an iterative procedure. In the following, we use this
coarse-grained tree-like structure and speak about trees, branches
etc. only in this sense. The cliques are numbered by
$\nu=1,...,(K+1)N/p$.  We closely follow the presentation and notation
of \cite{MePa}, cf. also \cite{BiMe}.

Let us consider a branch of the graph which is rooted in vertex $i$.
One of the cliques containing vertex $i$ is not part of the branch, we
denote it by $\nu$. We first calculate the partition function
$\Xi_{0/1}(i,\nu)$ of this branch where $i$ is fixed to be empty
($x_i=0$) or occupied ($x_i=1$). Let us denote the other $K$ cliques
containing $i$ by $\nu_1,...,\nu_K$, and the further vertices
contained in clique $\nu_k$ by $j_{1,k},...,j_{p-1,k}$, $k=1,...,K$.
Due to the hard-sphere constraint, every clique is allowed to carry at
most one particle, we thus find
\begin{eqnarray}
  \label{eq:Ziter}
  \Xi_0(i,\nu) &=& \prod_{k=1}^{K} \left[ \prod_{a=1}^{p-1 }
  \Xi_0(j_{a,k},\nu_k) + \sum_{a=1}^{p-1}\Xi_1(j_{a,k},\nu_k)
  \prod_{b\neq a}\Xi_0(j_{b,k},\nu_k) 
  \right] \nonumber\\
  \Xi_1(i,\nu) &=& e^\mu \prod_{k=1}^{K}
            \left[ \prod_{a=1}^{p-1 }\Xi_0(j_{a,k},\nu_k) \right] \ .
\end{eqnarray}
Introducing cavity fields $\mu h_i^\nu =
\ln(\Xi_1(i,\nu)/\Xi_0(i,\nu))$, these iterate according to
\begin{eqnarray}
  \label{eq:hiter}
  h_i^\nu &=& 1 + \sum_{k=1}^K 
  u(h_{j_{1,k}}^{\nu_k},...,h_{j_{p-1,k}}^{\nu_k})
  \nonumber\\
  u(h_1,...,h_{p-1}) &=& - \frac 1\mu \ln\left(
   1 + \sum_{a=1}^{p-1} e^{\mu h_a} \right)\ , 
\end{eqnarray}
i.e. they result from a linear superposition of the local chemical
potential $\mu$ and fields propagated along the cliques; and they take
values $h_i^\nu \in [1-K-\ln(p-1+e^{-\mu})/\mu,1]$. The true effective
field acting on a vertex is given by the contributions propagated from
all neighbors, i.e. from all $K+1$ cliques containing the vertex, $H_i
= 1 + \sum_{k=1}^{K+1}
u(h_{j_{1,k}}^{\nu_k},...,h_{j_{p-1,k}}^{\nu_k}) \in
[-K-\ln(p-1+e^{-\mu})/\mu,1]$.  Note that the r.h.s. of this equation
still contains the fields $h_j^\nu$ coming from single branches only.
The distribution of these fields serves as an order parameter and
allows to identify liquid, crystalline and glassy states.
Furthermore, its knowledge allows to calculate the grand canonical
potential \cite{MePa}, with ${\cal E}$ denoting the set of all
$p$-cliques in the graph, $ - \ln\Xi = - K \sum_i \ln\left( 1 + e^{\mu
    H_i}\right) + \sum_{\nu \in {\cal E} } \ln\left( 1 +
  \sum_{i\in\nu} e^{\mu h_i^\nu} \right), $ as well as the average
particle density $\rho$ per site, $ N \rho = {\overline{ \sum_i x_i}}
= \sum_i {e^{\mu H_i}}/{(1+e^{\mu H_i})} \ .  $

The liquid phase, which is expected to be globally stable for small
chemical potential $\mu$, is characterized by its translational
invariance. The effective fields $h_i^\nu$ and consequently $H_i$ do
not depend on the site index $i$ and are thus given by the homogeneous
fixed point of Eq. \ref{eq:hiter}.

For $K\geq 2$, this translational invariance breaks down in the
crystalline phase: On each clique, there is exactly one site of high
average density, all the other $(p-1)$ vertices have homogeneous lower
density. There are thus two fields $h^{(1)}<h^{(2)}$, which are a
solution of
\begin{eqnarray}
  \label{eq:crystall}
  h^{(1)} &=& 1 + K\ u(h^{(2)}, h^{(1)},..., h^{(1)}) \nonumber\\
  h^{(2)} &=& 1 + K\ u(h^{(1)},..., h^{(1)})\ .
\end{eqnarray}

For $p=2$, i.e. usual Bethe lattices, the crystalline solution appears
continuously due to a local instability of the liquid solution, the
solidification is thus of second order, and the model easily
crystallizes even if compactified relatively fast \cite{BeRi}.

In the remaining part of the paper we therefore concentrate to the
more interesting case $p\geq 3$, i.e. to graphs showing geometrical
frustration due to short loops. There the crystalline solution appears
discontinuously at a spinodal point $\mu_{\rm s}$. Still, its grand
canonical potential is larger than the one of the liquid phase.  The
two potentials cross at the crystallization transition $\mu_{\rm c}
>\mu_{\rm s}$, which thus describes a first-order transition being
accompanied by a discontinuous density jump. Sending $\mu\to\infty$,
the crystalline system quickly approaches the maximal density $1/p$.
For the example $p=4$, $K=3$, which is presented in Fig.
\ref{fig:tetrafixDensity}, the transition points are given by
$\mu_{\rm s}\simeq 1.73$ and $\mu_{\rm c}\simeq 2.18$.

The liquid phase stays, however, meta-stable even above $\mu_{\rm c}$
describing thus a super-cooled liquid. If we compactify the system
rather quickly, it initially stays liquid. However, at higher density,
the system even falls out of this local equilibrium: An exponential
number of glassy states of smaller density (and higher grand canonical
potential) appears discontinuously at a dynamical transition $\mu_{\rm
  d}$ and traps the system. At $\mu_{\rm rsb}$ a sub-exponential
number of glassy states finally reaches lower grand canonical
potential than the liquid one. In the equivalent replica approach,
this corresponds to one-step replica symmetry breaking.
For $p=4$ and $K=3$ we find $\mu_{\rm
  d}\simeq 5.7$ and $\mu_{\rm rsb}\simeq 6.1$, i.e. they are located
far beyond the crystallization transition; at $\mu_{\rm d}$ there are
about $\exp(0.0088 N)$ metastable states, but the clustering
phenomenon becomes more pronounced for larger $p$ and $K$.

Technically, this can be established using the cavity approach which
generalizes the Bethe-Peierls solution to more than one state
\cite{MePa}. The main difference to the liquid solution is that, in
different pure states, there are different cavity fields $h_i^\nu$.
They can be characterized by their histograms $P_i^\nu(h)$ 
which have to be determined self-consistently using the cavity
equations, for technical details see the analogous case discussed in
\cite{MePa}. With $P_i^\nu(h)$ being Dirac peaks in the liquid phase,
the dynamical transition point $\mu_{\rm d}$ can be identified by the
discontinuous appearance of a non-trivial solution for $P_i^\nu(h)$
which imidiately covers the full intervall of allowed fields. The
static transition $\mu_{\rm rsb}$ follows from the point where the
replica-symmetry broken grand canonical potential becomes dominant
with respect to the liquid one \cite{Mo2}.

For comparison, we have performed numerical simulations obtained on
fixed-connectivity random graphs built from 4-cliques. From an
analytical point of view they share the liquid and the glassy state
with the generalized Bethe lattices discussed above. Only the
crystalline state is inconsistent with the random character of the
graph entering via disordered loops of length ${\cal O}(N)$, and it
cannot be observed in the simulations. 
The simulation works as follows. For each MC step, a site $i$
is selected randomly. With probability $p=0.5$ a MOVE (M) step is
performed, and with probability $1-p$ an EXCHANGE (EX) step:
\begin{itemize}
\item[M] If site $i$ is empty and has exactly one occupied neighbor,
  the particle is moved to $i$. In all other cases, the configuration
  remains unchanged.
\item[EX] If the site is occupied, the particle is removed with
  probability $\exp(-\mu)$. If the site is empty, and all neighboring
  sites are empty, a particle is placed on $i$.
\end{itemize}
Note that in this way detailed balance is fulfilled, and the
equilibrium distribution is the one corresponding to Eq. (\ref{eq:xi}).
All simulations start with $\mu=0$. Then $\mu$ is increased in
steps of $\delta\mu=0.1$. For every value of the chemical potential,
we perform $n_{\rm MC}N$ Monte-Carlo steps.

In Fig. \ref{fig:tetrafixDensity}, the analytic findings are compared
with numerical results. Note that in principle the value of $\mu_{\rm
  d}$ can be inferred from the point where the numerical curve
deviates from the liquid one in the limit of infinitesimal slow
compaction ($n_{\rm MC}\to \infty$). Even if our data turned out to be
incompatible with every tested form of fits (as a function of $n_{\rm
  MC}$), this point appears to be consistent with the analytical
prediction which is marked by a vertical line in the figure.

As already mentioned, also the generalized Bethe lattices are or
mean-field nature. Whereas a similar glassy behavior was found also
numerically in a two-dimensional model of cross-shaped particles
\cite{EiBa}, it is useful to test our theoretical predictions at least
qualitatively in a finite-dimensional model which is closer to our
mean-field model. In particular it should be characterized by the same
kind of geometrical frustration.  We thus consider a three-dimensional
lattice with $N=L^3$ sites. It is generated from a tetrahedron, i.e. a
4-clique, which is translated along its edges.  Again each site is
contained in $K+1=4$ tetrahedra.  We impose periodic boundary
conditions.  On top of this lattice, we again consider our lattice gas
of hard spheres of radius one, and we numerically study its behavior
by means of MC simulations up to size $N=100^3$.

As in its mean-field counter-part, the system is found to be in a
liquid phase for small $\mu$: All particles are free to move around,
they are not localized, and the time-averaged single-site density is
homogeneous. For slow compaction, we observe a discontinuous
crystallization transition at $\mu\simeq2$, cf.\ Fig.\ 
\ref{fig:cliqueDens}. At this point, the symmetry between sites breaks
down, every tetrahedron has one site of higher than average density,
and three of lower density.  The average density exhibits a
discontinuity and quickly approaches 0.25, i.e. its theoretically
maximal value, if we further increase $\mu$. The first order character
of this transition can be numerically tested by decreasing $\mu$
again. Doing this we observe a hysteresis, cf.\ Fig.\ 
\ref{fig:cliqueDens}, the system stays in the crystalline state even
for $\mu < 2$.

If, starting from the liquid state at low $\mu$, we increase $\mu$
quite fast, crystallization is avoided. The system stays at lower than
the crystalline density. This phenomenon appears in analogy to the
super-cooled, meta-stable liquid observed in the mean-field model.  At
higher $\mu$, the system starts to freeze into a glassy state, being
characterized by a density which is much smaller than 0.25.  This
asymptotic density depends heavily on the compaction rate.

\begin{figure}
\twofigures[width=0.42\textwidth]{clique_dens_hyst.eps}{clique_wait2.eps}
\caption{ 
  Density $\rho$ of the lattice model
as a function of the chemical potential $\mu$ for
  $n_{\rm MC}=1,2,5,25,1000$ Monte-Carlo sweeps per value of the
  chemical potential (from bottom to top). All $n_{\rm MC}$ sweeps,
  the chemical potential is increased by 0.1. The data are for one run
  of a lattice with $N=50^3$ sites. The solid line represents the
  analytical result for the crystalline phase on the generalized Bethe
  lattice which can be considered as a simple mean-field approximation.
  Inset: Density $\rho$ as a function of $\mu$ for slowly increasing
  ($n_{\rm MC}=500$) and subsequently fast decreasing ($n_{\rm MC}=5$)
  chemical potential.}
\label{fig:cliqueDens}
\caption{
  Density $\rho$ of the lattice model as a function of the chemical
  potential $\mu$ for $n_{\rm MC}=2$, when the chemical potential is
  first decreased every $n_{\rm MC}$ sweeps by 0.1 until $\mu_{\rm
    stop}$ is reached. Then the system continues to evolve at fixed
  value of $\mu$. The inset shows the density as a function of time
  for smaller $\mu_{\rm stop}$. The arrows indicate the times from
  where the chemical potential is kept fixed. The horizontal lines
denote the equilibrium densities of the crystalline state at $\mu_{\rm
  stop}$.}
\label{fig:cliqueWait}
\end{figure}

There are, however, some crucial differences between Figs.
\ref{fig:tetrafixDensity} and \ref{fig:cliqueDens}. If we consider the
compaction curves in the mean-field model, cf. Fig.
\ref{fig:tetrafixDensity}, they overlap even beyond the
crystallization transition, until the (local) equilibration time
start to be larger than the characteristic compaction time, and the
system fall out of the meta-stable equilibrium. If we stop to increase
$\mu$ at some point below the glass transition, the density evolves
towards the one of the liquid phase, i.e. the latter is a true
metastable phase. This seems not to be true for the finite-dimensional
lattice. The compaction rate dependence sets in directly at the
crystallization transition, and there is no pronounced gap between a
meta-stable liquid state and the globally stable solid one. In Fig.
\ref{fig:cliqueWait} we report the following experiment: The system is
compactified quite fast by increasing the chemical potential up to
some value $\mu_{\rm stop}$ which is chosen beyond the crystallization
point. Then the system is allowed to evolve at fixed $\mu$. As can be
seen in the inset for a system of size $N=50^3$, the model still
equilibrates towards the crystalline state (given by the horizontal
lines), even if this equilibration time is much larger than the
equilibration time in the liquid phase. In the main part of
Fig. \ref{fig:cliqueWait}, we show the system size dependence of this
effect: The equilibration time grows with $N$, and probably diverges
in the thermodynamic limit. This growth is not exponential in $N$,
i.e. there are no extensive energy barriers to be overcome as in the
generalized Bethe-lattice model.

To conclude, we have introduced a hard-core lattice-gas model with
only two-particle interactions.  On a generalized Bethe lattice, it
exhibits a crystalline phase as well as glassy behavior if
super-cooled sufficiently fast. In contrast to usual Bethe lattices,
the generalized ones have short loops of odd length. These loops, and
the resulting geometrical frustration, are responsible for the glassy
behavior of the model which is similar to the behavior known from
models with multi-spin interactions: If super-cooled, the system falls
out of equilibrium at a dynamical transition which is accompanied by
the appearance of an exponential number of states. Also in our model,
this transition is not of thermodynamic nature, a replica-symmetry
breaking equilibrium transition occurs only at higher chemical
potential.  It is quite interesting that similar phenomena can also be
found in a completely different model considered in \cite{Nu}: There,
geometrical frustration is introduced into a field theoretical setting
in form of a small non-Abelian background gauge field. In a sense,
these different approaches complement and re-enforce each other.

Furthermore, we have studied numerically a three-dimensional variant
of the model exhibiting the same local parameters, like sizes of the
short loops and number of neighbors. The behavior of the lattice model
is close to the mean-field type system except that the lattice variant
does not have a true metastable phase which remains stable at the
timescale of the simulation.

Another conclusion concerns the typical numerical hardness
of the vertex-cover problem, which is equivalent to finding dense
hard-sphere packings \cite{WeHa}: Graphs containing small loops, or
geometrical frustration, are expected to be much harder to cover with
a small vertex cover size than locally tree-like graphs like random
graphs. This opens the interesting question in how far local
structures may influence the global hardness of a combinatorial
problem.

\acknowledgments
 
We acknowledge interesting an fruitful discussions with J. Berg, F.
Ricci-Tersenghi, and R. Zecchina. AKH obtained financial support from
the DFG (Deutsche Forschungsgemeinschaft) under grants Ha 3169/1-1 and
Zi 209/6-1. MW thanks the ICTP Trieste for hospitality.

\end{document}